\documentclass[aps,prc,twocolumn,superscriptaddress,secnumarabic]{revtex4}
\usepackage{amsmath}   
\usepackage{amsfonts}  
\usepackage{graphicx}  
\usepackage{verbatim}  
\usepackage{color}     
\usepackage{array}
\usepackage{subfigure} 
\usepackage{hyperref}  
\usepackage{breakurl}

\begin{document}

\newcommand{\gray}{\ensuremath{\gamma}-ray}
\newcommand{\grays}{\ensuremath{\gamma}-rays}
\newcommand{\vgray}{\ensuremath{\vec{\gamma}}-ray}
\newcommand{\vgrays}{\ensuremath{\vec{\gamma}}-rays}
\newcommand{\vgamma}{\ensuremath{\vec{\gamma}}}
\newcommand{\HIGS}{HI\vgamma S}
\newcommand{\etal}         {{\em et~al.}}
\newcommand{\ie}         {{\em i.e.}}
\newcommand{\eg}         {{\em e.g.}}
\newcommand{\nuc}[2]{\ensuremath{^{#1}}#2}
\newcommand{\geant} {{\bf {G}\texttt{\scriptsize{EANT}}4}}

\newcommand{\rfig}[1]       {Figure \ref{fig:#1}}
\newcommand{\rfigs}[3]       {Figures \ref{fig:#1}, \ref{fig:#2}, and \ref{fig:#3}}
\newcommand{\req}[1]       {Eqn. \ref{eq:#1}}
\newcommand{\rtab}[1]       {Table \ref{tab:#1}}

\title{Compton Scattering from \nuc{6}{Li} at 86 MeV}

\author{L.S. Myers}
  \altaffiliation{Present address: Thomas Jefferson National Accelerator Facility, Newport News, VA 23606, USA}
\affiliation{Department of Physics, Duke University, Durham, NC 27708, USA}
\affiliation{Triangle Universities Nuclear Laboratory, Durham, NC 27708, USA}

\author{M.W. Ahmed}
\affiliation{Department of Physics, Duke University, Durham, NC 27708, USA}
\affiliation{Triangle Universities Nuclear Laboratory, Durham, NC 27708, USA}
\affiliation{Department of Mathematics and Physics, North Carolina Central University, Durham, NC 27707, USA}

\author{G. Feldman}
\affiliation{Department of Physics, George Washington University, Washington, DC 20052, USA}

\author{A. Kafkarkou}
  \altaffiliation{Deceased.}
\affiliation{Department of Physics, Duke University, Durham, NC 27708, USA}
\affiliation{Triangle Universities Nuclear Laboratory, Durham, NC 27708, USA}

\author{D.P. Kendellen}
\affiliation{Department of Physics, Duke University, Durham, NC 27708, USA}
\affiliation{Triangle Universities Nuclear Laboratory, Durham, NC 27708, USA}

\author{I. Mazumdar}
\affiliation{Tata Institute of Fundamental Research, Colaba, Mumbai, 400 005, India}

\author{J.M. Mueller}
\affiliation{Department of Physics, Duke University, Durham, NC 27708, USA}
\affiliation{Triangle Universities Nuclear Laboratory, Durham, NC 27708, USA}

\author{M.H. Sikora}
\affiliation{Triangle Universities Nuclear Laboratory, Durham, NC 27708, USA}
\affiliation{Department of Physics, George Washington University, Washington, DC 20052, USA}

\author{H.R. Weller}
\affiliation{Department of Physics, Duke University, Durham, NC 27708, USA}
\affiliation{Triangle Universities Nuclear Laboratory, Durham, NC 27708, USA}

\author{W.R. Zimmerman}
\affiliation{Triangle Universities Nuclear Laboratory, Durham, NC 27708, USA}

\date{\today}

\begin{abstract}

Cross sections for \nuc{6}{Li}($\gamma$,$\gamma$)\nuc{6}{Li} have been measured at the High Intensity Gamma-Ray Source (\HIGS) and the sensitivity of these cross sections to the nucleon isoscalar polarizabilities was studied. Data were collected using a quasi-monoenergetic 86~MeV photon beam at photon scattering angles of 40$^{\circ}$--160$^{\circ}$. These results are an extension of a previous measurement at a lower energy. The earlier work indicated that the \nuc{6}{Li}($\gamma$,$\gamma$)\nuc{6}{Li} reaction at 60~MeV provides a means of extracting the nucleon polarizabilities; this work demonstrates that the sensitivity of the cross section to the polarizabilities is increased at 86~MeV. A full theoretical treatment is needed to verify this conclusion and produce values of the polarizabilities.


\end{abstract}

\maketitle

\section{Introduction}

Compton scattering from nucleons and light nuclei are particularly sensitive to the nucleon polarizabilities, which are fundamental structure constants relating the nucleon response to an applied electric or magnetic field. Cross sections for \nuc{6}{Li}($\gamma$,$\gamma$)\nuc{6}{Li} and \nuc{16}{O}($\gamma$,$\gamma$)\nuc{16}{O} at E$_\gamma$~=~60~MeV measured at the High Intensity Gamma-Ray Source (\HIGS) were reported in a previous paper \cite{Myers2012} to study the feasibility of using this technique on \nuc{6}{Li}. The conclusion of that paper was that, given a detailed theoretical treatment, measurements of Compton scattering cross sections from \nuc{6}{Li} at the level of accuracy available at the \HIGS\ facility will improve our knowledge of the nucleon isoscalar electric ($\alpha$) and magnetic ($\beta$) polarizabilities.

The advantage of using \nuc{6}{Li} as a target in nuclear Compton-scattering experiments is that its cross section is nearly an order of magnitude larger than that of hydrogen and deuterium -- the most commonly used targets for studying $\alpha$ and $\beta$. The larger cross section of \nuc{6}{Li} reduces the statistical uncertainty associated with the measurement at the expense of having a more complicated nucleus to model. The success of the first measurement was the motivation for repeating the experiment at a higher photon energy. Although the Compton-scattering cross section tends to decrease as a function of energy, the sensitivity to the polarizabilities increases with higher energies. For that reason, a new measurement of \nuc{6}{Li}($\gamma$,$\gamma$)\nuc{6}{Li} has recently been performed at E$_\gamma$~=~86~MeV.

In this paper, we will briefly review the experimental setup used to measure Compton-scattering cross sections at \HIGS\ and then report results for both \nuc{16}{O}($\gamma$,$\gamma$)\nuc{16}{O} and \nuc{6}{Li}($\gamma$,$\gamma$)\nuc{6}{Li} at E$_\gamma$~=~86~MeV. At the time of this publication, a full theoretical calculation for the measured \nuc{6}{Li} cross sections has not yet been developed, so only a brief discussion of the possibility of extracting values of $\alpha$ and $\beta$ from these data will be presented in this paper.

\section{Experimental Setup}

We present measurements of the \nuc{16}{O}($\gamma$,$\gamma$)\nuc{16}{O} and \nuc{6}{Li}($\gamma$,$\gamma$)\nuc{6}{Li} Compton-scattering cross sections at a photon energy of 86~MeV. The \HIGS\ facility \cite{Litvinenko97,Weller09} is capable of producing $\sim$100$\%$ circularly-polarized, quasi-monoenergetic $\gamma$-ray beams. Experiments at energies of up to $\sim$100~MeV are now feasible because of recent upgrades to the \HIGS\ facility \cite{Wu2013}. Since the \nuc{16}{O} cross section is well-known at 86~MeV from previous measurements \cite{Feldman96}, we have repeated the \nuc{16}{O} measurement at this energy in order to establish the systematics of the absolute cross-section extraction for the higher-energy \nuc{6}{Li} data. 

\begin{figure}[htb]
 \begin{center}
 \includegraphics[width=\columnwidth]{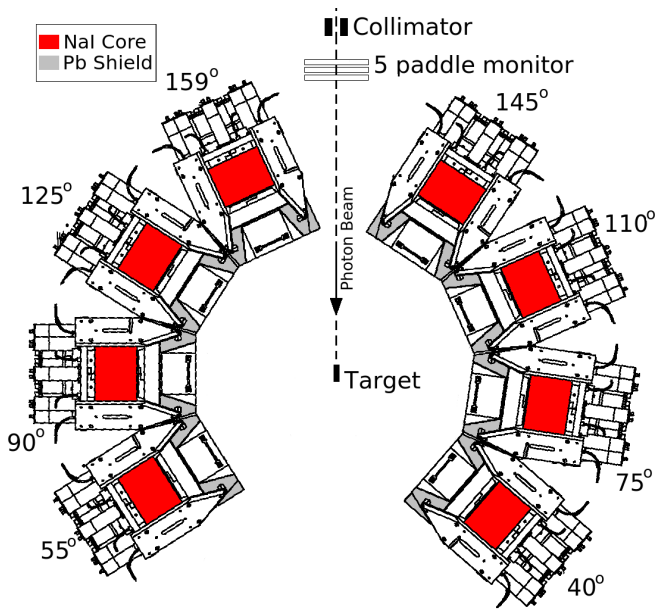}
 \caption{(Color online) Diagram of the experimental layout showing the positions of the collimator, five-scintillator-paddle system, target, and NaI(Tl) detectors. Drawing is not to scale.}
 \label{fig:Layout}
 \end{center}
\end{figure}

The \HIGS\ facility produces $\gamma$-ray beams through Compton backscattering using a free-electron laser. The beam is collimated to define the beam-spot diameter. Collimation also determines the intensity and energy spread of the incident $\gamma$ rays. For this experiment, with a 14 mm diameter collimator, the 86~MeV \gray\ beam had an on-target intensity of $\sim$4 $\times$ 10$^6$~$\gamma$/s and an energy spread of $\sim$6$\%$ \cite{Mikhailov13}. A five-scintillator-paddle system \cite{Pywell09} was used to monitor the $\gamma$-ray beam intensity. After the beam passed the intensity monitor, the $\gamma$ rays impinged upon the scattering target (\nuc{6}{Li} or \nuc{16}{O}). Eight NaI(Tl) detectors placed at roughly equally spaced angles from 40$^{\circ}$--160$^{\circ}$, as shown in \rfig{Layout}, were used to detect the scattered $\gamma$ rays.

Each NaI(Tl) detector assembly is composed of a cylindrical core of 25.4~cm diameter and 30.5~cm length, snugly fitted inside a segmented NaI(Tl) annulus of 30.5~cm length and 7.5~cm thickness. The annular detector was designed as an anti-coincidence shield but was not used during this experiment. Each detector had a lead shield placed at its front face to reduce backgrounds and define the scattering solid angle ($\sim$45~msr at a distance of 58.5~cm from the target). The conical aperture in the lead shielding was filled with borated wax to significantly reduce neutron backgrounds without any significant attenuation of the incident photon flux on the face of the detector and minimally affecting the resolution.

The oxygen target consisted of distilled water in a Lucite cylinder 11.3~cm long and 5.0~cm in diameter, giving a target thickness of 11.3$\pm$0.1~g/cm$^2$. The contribution to the scattering yield from the target end caps was estimated to be less than 1$\%$. The \nuc{6}{Li} target, produced at the University of Saskatchewan \cite{Wurtz07}, consisted of a polyvinyl chloride (PVC) cylinder with PVC and Al foil end caps. It was 12.7~cm long and 4.1~cm in diameter for a target thickness of 5.84$\pm$0.06~g/cm$^2$. An identical, empty-target cell was also provided. The contribution from the cell end caps was estimated to be $\sim$2$\%$ so a measurement from the empty cell was taken as well. The empty-target yields were then subtracted from the full-target yields to obtain the scattering yield from \nuc{6}{Li}.

\section{Data Analysis}

\begin{figure}[htb]
 \begin{center}
 \includegraphics[width=\columnwidth]{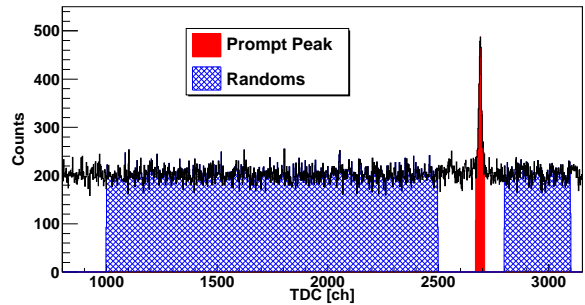}
 \caption{(Color online) A representative TDC spectrum with the prompt and random windows shown.}
 \label{fig:TDC}
 \end{center}
\end{figure}

The signals from the NaI(Tl) detectors were read into a charge-to-digital converter (QDC) and a time-to-digital converter (TDC). The QDC channels were calibrated into equivalent photon energy using the elastically-scattered photons whose energies are known from kinematics. The pulsed nature of the photon beam at \HIGS\ allows coincident timing in the TDC spectra (see \rfig{TDC}). The TDC is started by the signal from a scattered photon in the NaI and stopped by a capacitive pickoff signal from the beam pulse. The coincidence peak in the TDC spectrum corresponds to beam-related $\gamma$ rays that are detected in the NaI.

\begin{figure}[htb]
 \begin{center}
 \includegraphics[width=\columnwidth]{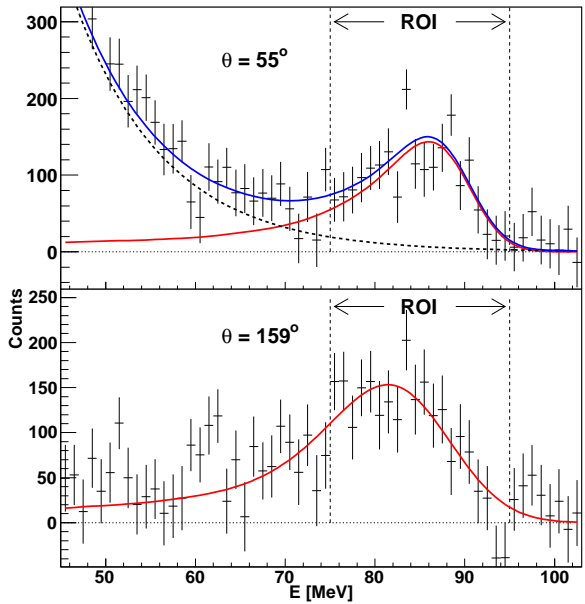}
 \caption{(Color online) The energy spectrum (with both random and empty-target contributions subtracted) at forward and backward angles for \nuc{6}{Li}. The dashed, vertical lines denote the region of interest (ROI). The red curves indicate the elastic peak determined by fitting the \geant\ simulations. At forward angles, the atomic-related background is modeled by an exponential (dashed curve), and the total response is shown by the blue curve.}
 \label{fig:QDCLi}
 \end{center}
\end{figure}

QDC spectra for prompt and random events were generated by placing cuts on the prompt and random windows in the TDC spectrum. The prompt window contains both true scattered photons and some fraction of random events. Subtraction of the random contribution produces the net true spectrum, defined as 
\begin{equation}\label{eq:subtraction}
S_{net} = S_{prompt} - \frac{W_{prompt}}{W_{random}} \times S_{random},
\end{equation}

\noindent where $W_{prompt}$ and $W_{random}$ represent the widths of the prompt and random windows, and $S_{net}$, $S_{prompt}$ and $S_{random}$ are the net, prompt and random energy spectra. The empty-target data were analyzed in the same manner as the full target, then subtracted from the full-target spectrum after being normalized to the incident photon intensity. The yield of the empty target, normalized to the number of incident photons, in the elastic peak region (indicated by the region of interest (ROI) in \rfig{QDCLi}) was $<3\%$ of the yield in the full target. Typical final detector-response line shapes are shown for \nuc{6}{Li} (\rfig{QDCLi}) at both forward and backward scattering angles. The low-energy background in the forward-angle detector is assumed to be coming from forward-peaked, atomic-scattering events in the target. The backward-angle detectors do not see this background. 

Simulations using \geant\ (version 4.9.2) \cite{geant} were used to model the elastic-scattering line shape of the detector and to correct for any effects resulting from the finite geometry of the experimental setup. As shown previously in \cite{Myers2012}, \geant\ is able to accurately reproduce the measured response function of the NaI(Tl) detectors when the simulated response is convoluted with a Gaussian distribution. The parameters of the convolution are consistent between the oxygen and lithium data sets despite the significant difference in the signal-to-noise ratio. The low-energy background at $\theta_{\rm Lab}$ $<$ 90$^\circ$ was modeled by an exponential curve; the sum of the curve and the simulated elastic peak was then fit to the data to obtain the total response of the detector. Variation of the exponential background indicated that the random uncertainty introduced is $\sim$5\%. The total photon yield was extracted from the fitted line shapes and normalized to the number of photons on target, the target thickness, and the solid angle to obtain the scattering cross section. In the final data set, only 7 of the 8 NaI detectors were included as the detector at $\theta_{\rm Lab}$~=~125$^\circ$ displayed significant and uncorrectable gain shifts and was removed from the analysis.

\begin{figure}[htb]
 \begin{center}
 \includegraphics[width=\columnwidth]{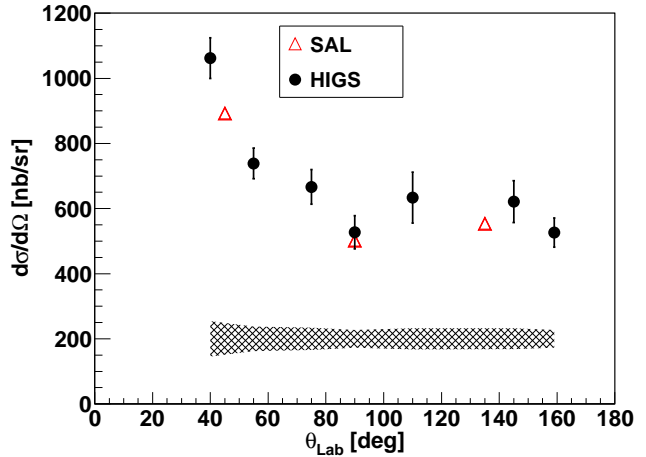}
 \caption{(Color online) Comparison of scattering cross section for \nuc{16}{O} obtained by \cite{Feldman96} and present results. The statistical uncertainties are shown on the present results and the systematic uncertainties for this measurement are shown in the band beneath the data.}
 \label{fig:OxygenPlot}
 \end{center}
\end{figure}

The cross sections for \nuc{16}{O} from the current experiment and from the Saskatchewan Accelerator Laboratory (SAL) \cite{Feldman96} are shown in \rfig{OxygenPlot}. The agreement is very good between the two data sets. These data would support the conclusion of \cite{Feldman96} that the nucleon polarizabilities suffer some modification when the nucleons are bound within a nucleus.

The extracted cross sections for \nuc{16}{O} and \nuc{6}{Li} are given in \rtab{CrossSections} along with the statistical and systematic uncertainties. The statistical uncertainty includes the random uncertainties (added in quadrature) arising from the extraction of the photon yield and the modeling of the low-energy background. The systematic errors are dominated by the uncertainty in the number of incident photons (3.0$\%$) and the detector acceptance (3.7$\%$). 

Based on the analysis of the E$_\gamma$~=~60~MeV data \cite{Myers2012}, any contribution from inelastic scattering was expected to be less than or of the order of a few percent of the total photon yield. The poorer resolution of the elastic peak in this experiment made it more difficult to estimate the potential contribution from inelastic scattering. By fitting the back-angle detectors with the sum of elastic and inelastic scattering spectra the potential contribution of inelastic scattering was $\sim$1--25\% of the elastic yield, but the contribution at each angle was also consistent with zero within uncertainties. Since we can not confirm the presence of significant inelastic contribution, the cross sections and uncertainties reported here assume that inelastic scattering can be ignored.

\begin{table}
  \setlength{\extrarowheight}{6pt}
  \begin{center}
    \caption{
      Compton scattering cross sections for \nuc{16}{O} and \nuc{6}{Li} at $E_{\gamma}$~=~86~MeV. The first uncertainty is statistical and the second is systematic.
    }
    \begin{tabular*}{0.75\columnwidth}{@{\extracolsep{\fill}} c c c }
      \hline
      \hline
       & \nuc{16}{O} & \nuc{6}{Li} \\
      $\theta_{\rm Lab}$ & $\frac{d\sigma}{d\Omega}$ (nb/sr) & $\frac{d\sigma}{d\Omega}$ (nb/sr) \\
      \hline
      40$^{\circ}$ & 1062 $\pm$ 63 $\pm$ 58 & 203 $\pm$ 18 $\pm$ 10 \\
      55$^{\circ}$ & \ 739 $\pm$ 47 $\pm$ 38 & 147 $\pm$ 11 $\pm$ \ 7 \\
      75$^{\circ}$ & \ 667 $\pm$ 53 $\pm$ 34 & 140 $\pm$ 12 $\pm$ \ 7 \\
      90$^{\circ}$ & \ 527 $\pm$ 51 $\pm$ 27 & 160 $\pm$ 13 $\pm$ \ 8 \\
      110$^{\circ}$ & \ 634 $\pm$ 78 $\pm$ 33 & 146 $\pm$ 20 $\pm$ \ 7 \\
      145$^{\circ}$ & \ 622 $\pm$ 64 $\pm$ 32 & 167 $\pm$ 18 $\pm$ \ 8 \\
      159$^{\circ}$ & \ 527 $\pm$ 45 $\pm$ 27 & 172 $\pm$ 12 $\pm$ \ 9 \\
      \hline
      \hline
    \end{tabular*}
    \label{tab:CrossSections}
  \end{center}
\end{table}

\section{Results}

\begin{figure}[htb]
 \begin{center}
 \includegraphics[width=\columnwidth]{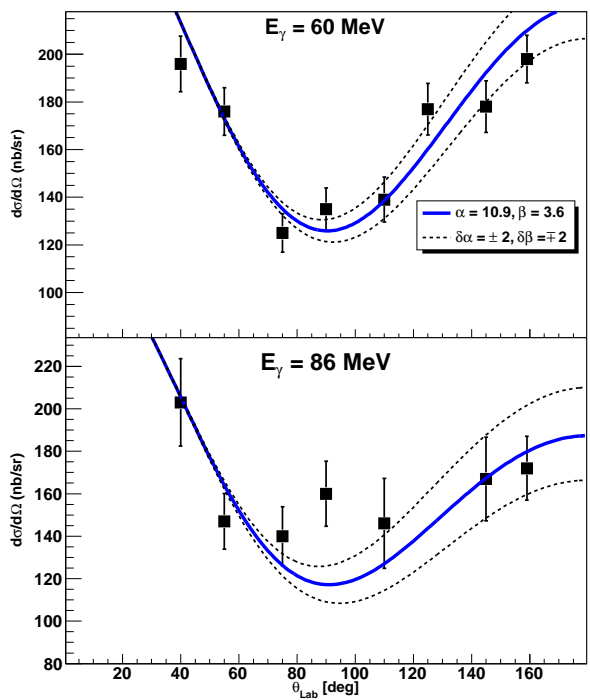}
 \caption{(Color online) The \nuc{6}{Li} Compton-scattering cross section from this experiment at 86~MeV (bottom) and measured previously at 60~MeV (top). The error bars shown on the data points are the statistical and systematic uncertainties added in quadrature. The curves are the results obtained from fitting the data to a phenomenological model \cite{Myers2012}.}
 \label{fig:LithiumPlot}
 \end{center}
\end{figure}

The \nuc{6}{Li} cross sections are plotted in \rfig{LithiumPlot} (bottom panel) along with the earlier \HIGS\ data from 60~MeV (top panel). We have combined the earlier 60~MeV data with the current higher-energy data for the purpose of performing a comprehensive analysis to attempt to determine the sensitivity to the nucleon polarizabilities. The complete \nuc{6}{Li}($\gamma$,$\gamma$)\nuc{6}{Li} data set was evaluated using the phenomenological model detailed in \cite{Myers2012,Feldman96,Wright85}. This model uses as inputs the isovector E1 and E2 giant resonances, the quasi-deuteron process (QD), and the isoscalar polarizabilities. It is not possible to extract values of $\alpha$ and $\beta$ using this model, in part, because the giant resonance states are not well-known. However, the model should suffice to indicate the degree to which the data are sensitive to the polarizabilities.

The QD and giant resonances are modeled as Lorentzians of the form
\begin{equation}\label{eq:Imf}
  \sigma_\lambda(E) = \frac{\sigma^0_\lambda E^2 \Gamma_\lambda^2}{(E^2-E_\lambda^2)^2 + E^2\Gamma_\lambda^2}
\end{equation}

\noindent where $E_\lambda$, $\Gamma_\lambda$ and $\sigma_\lambda$ are the resonance energy, width, and strength, respectively, of the QD or the E1 or E2 resonance. By making some basic assumptions about the location and width of these resonances, and fixing $\alpha$ and $\beta$ to previous values \cite{Griesshammer12}, the strength of the resonances can be extracted by fitting the data sets with the model. The resonance parameters found through this approach are given in \rtab{ResPars}. These parameters are nearly identical to those reported in \cite{Myers2012}.

\begin{table}
  \setlength{\extrarowheight}{6pt}
  \begin{center}
    \caption{
      E1, E2, and QD resonance parameters for \nuc{6}{Li} determined from fitting.
    }
    \begin{tabular*}{0.75\columnwidth}{@{\extracolsep{\fill}} c c c c }
      \hline
      \hline
      Resonance & $E_{res}$ & $\Gamma_{res}$ & $\sigma_{res}$ \\
       & (MeV) & (MeV) & (mb) \\
      \hline
      E1 & 25.0 & 12.0 & \ 7.6 \\
      E2 & 34.0 & 16.0 & 0.20 \\
      QD & 40.0 & 100 & \ 1.1 \\
      \hline
      \hline
    \end{tabular*}
    \label{tab:ResPars}
  \end{center}
\end{table}

The resonance parameters were fixed to the values in \rtab{ResPars} in order to determine the sensitivity of the data to $\alpha$ and $\beta$. Initially, the polarizabilities were varied by $\pm$2 $\times$ 10$^{-3}$~fm$^3$ (see \rfig{LithiumPlot}) while imposing the Baldin Sum Rule \cite{Baldin60}, which constrains $\alpha$+$\beta$~=~14.5 $\times$ 10$^{-3}$~fm$^3$. The dotted curves represent this range of values. Numerically, a fit of the model to the data (using the combined statistical and systematic uncertainties added in quadrature) produces an uncertainty in $\alpha$ ($\delta \alpha$~=~0.7 $\times$ 10$^{-3}$~fm$^3$) that compares very favorably to the value obtained from deuterium Compton scattering ($\delta \alpha$~=~0.9 $\times$ 10$^{-3}$~fm$^3$, but only considering statistical uncertainties) \cite{Griesshammer12}. This result would appear to support the initial premise that Compton scattering from \nuc{6}{Li} provides a new avenue for extracting the nucleon polarizabilities. The statistical and systematic uncertainties should be comparable to or better than those obtained in deuteron Compton scattering, primarily due to the larger \nuc{6}{Li} cross section. Additionally, \nuc{6}{Li} is a simpler target to utilize experimentally since it does not require the cryogenics used to maintain liquid hydrogen or deuterium targets. This work, together with a proper theoretical treatment, indicates that \nuc{6}{Li} may be a preferred target for high precision extraction of the nucleon polarizabilities. 

The limitations of the present model and the parameters needed to apply it prevent an extraction of $\alpha$ and/or $\beta$ from these data.  However, our analysis does indicate that Compton-scattering data from \nuc{6}{Li} in the vicinity of 86~MeV exhibit a strong sensitivity to the nucleon isoscalar polarizabilities. An accurate theoretical treatment of the reaction cross section is needed in order to verify this conclusion. The results of this experiment are intended to motivate future work on theoretical models of Compton scattering from light nuclei and \nuc{6}{Li} in particular.

The authors would like to acknowledge the contributions of the staff at the \HIGS\ facility for the production of the \gray\ beam used in this experiment. This work was supported in part by the U.S. Department of Energy, Office of Science Grant Nos. DE-FG02-97ER41033 and DE-FG02-06ER41422.

\end{document}